\documentclass[aps,prl,superscriptaddress,twocolumn,twoside]{revtex4}
\pdfoutput=1

\usepackage[utf8]{inputenc}

\usepackage{mathrsfs,mathtools,bbold}
\usepackage{amsmath,amssymb}
\usepackage{amsfonts}
\usepackage{verbatim}
\usepackage{graphicx}
\usepackage{mathrsfs,mathtools}
\usepackage{ccaption}
\usepackage[dvipsnames]{xcolor}
\definecolor{SchoolColor}{rgb}{0.6471, 0.1098, 0.1882} 
\usepackage{stmaryrd}
\usepackage{textcomp}
\usepackage{eucal} 
\usepackage{setspace}
\usepackage{empheq}
\usepackage[colorlinks=true,linkcolor=black,citecolor=teal,urlcolor=MidnightBlue,filecolor=black]{hyperref}

\DeclareMathOperator{\extdm}{d}
\newcommand{\extd}{\extdm \!}

\providecommand{\Lt}{{\tt L}}
\renewcommand{\Lt}{{\tt L}}
\providecommand{\Mt}{{\tt Mt}}
\renewcommand{\Mt}{{\tt M}}

\newcommand{\eq}[2]{\begin{equation} #1 \label{#2} \end{equation}}   
   
\begin{document}

\title{Local quantum energy conditions in non-Lorentz-invariant quantum field theories}

\author{Daniel Grumiller}
\email[]{grumil@hep.itp.tuwien.ac.at}

\affiliation{Institute for Theoretical Physics, TU Wien, Wiedner Hauptstr. 8, A-1040 Vienna, Austria}

\author{Pulastya Parekh}
\email[]{pulastya@iitk.ac.in}

\affiliation{Erwin Schr$\ddot{o}$dinger International Institute for Mathematics and Physics, 1090 Vienna, Austria.}
\affiliation{Indian Institute of Technology Kanpur, Kalyanpur, Kanpur 208016,  India}

\author{Max Riegler}
\email[]{max.riegler@ulb.ac.be}

\affiliation{Universit\'{e} Libre de Bruxelles and International Solvay Institutes, CP 231, B-1050 Brussels, Belgium}

\date{\today}

\begin{abstract}
We provide the first example of local quantum energy conditions in quantum field theories that are not Lorentz invariant. We focus on field theories in two dimensions with infinite-dimensional symmetries, like the ones governed by the Bondi-van der Burg- Metzner-Sachs group that appear in the context of flat space holography. Reminiscent of holographic results on the quantum null energy condition, we prove that our new energy conditions saturate for states in the field theory that are dual to vacuum solutions of three-dimensional Einstein gravity with vanishing cosmological constant.
\end{abstract}

\maketitle

\hypersetup{linkcolor=SchoolColor}



\paragraph{Introduction.} Among the various local energy conditions the Quantum Null Energy Condition (QNEC) \cite{Bousso:2015mna} has attracted considerable amount of interest over recent years \cite{Fu:2016avb,Fu:2017evt,Akers:2017ttv,Ecker:2017jdw,Fu:2017ifb,Leichenauer:2018obf,Khandker:2018xls,Ecker:2019ocp} including various proofs \cite{Bousso:2015wca,Koeller:2015qmn,Balakrishnan:2017bjg}. This is due to the fact that given certain assumptions, notably unitarity, QNEC can be shown to hold universally for quantum field theories (QFTs) in dimensions $D\geq2$ \cite{Ceyhan:2018zfg}, unlike any other local energy condition. For the special case of two-dimensional conformal field theories (CFT$_2$) QNEC takes a particularly simple form
    \begin{equation}\label{eq:AdSQNEC}
        2\pi\left\langle\mathcal{T_{\pm\pm}}\right\rangle\geq S''+\frac{6}{c}\,S'^2
    \end{equation}
where $\left\langle\mathcal{T_{\pm\pm}}\right\rangle$ denotes expectation values of null projections of the energy-momentum tensor for a given state, $c>0$ is the central charge of the unitary CFT$_2$, and $S$ is the entanglement entropy (EE) of an arbitrary interval where one of the endpoints coincides with the locus at which the stress tensor is evaluated. Primes denote variations with respect to deformations of this endpoint into the same null direction as used for the projection on the left hand side of \eqref{eq:AdSQNEC}.

Lorentz invariance plays an important role for QNEC, as the appearance of the word `null' suggests. However, there are QFTs that do not exhibit Lorentz invariance, for example the worldsheet theory of tensionless strings \cite{Gamboa:1989px,Isberg:1993av,Bagchi:2013bga,Bagchi:2015nca}, Galilean \cite{Bagchi:2009my,Bagchi:2017yvj} or warped conformal field theories \cite{Detournay:2012pc} and hence the very notion of `null' need not exist in non-Lorentz-invariant QFTs. This means that QNEC cannot be universally true for such QFTs. It is interesting to investigate whether inequalities analogous to (but different from) QNEC hold for such theories. To distinguish such energy conditions from the specific case of QNEC we shall refer to them generally as ``Quantum Energy Conditions'' (QECs). 

The main goal of the present work is to establish the first example of such a QEC for a specific class of non-Lorentz-invariant field theories. One of the main tools we shall use is holography, which we briefly review now.

\paragraph{Holography and flat space limit.} The holographic principle has played a vital role in deepening our understanding of quantum gravity. A particularly famous avatar of such a correspondence is the celebrated Anti-de~Sitter/conformal field theory (AdS/CFT) correspondence that relates type IIB superstring theory on AdS$_5\otimes S^5$ to $\mathcal{N}= 4$ supersymmetric Yang-Mills theory on its boundary \cite{Maldacena:1997re}. Roughly a decade after the dawn of AdS/CFT, Ryu and Takayanagi provided a holographic interpretation of entanglement entropy (EE) \cite{Ryu:2006bv,Ryu:2006ef} (see \cite{Hubeny:2007xt} for a covariant formulation) showing a deep relation between gravity and quantum information.

While AdS/CFT is an immensely successful example of a holographic correspondence there have been intense efforts to find other instances where the holographic principle \cite{'tHooft:1993gx,Susskind:1995vu} is realized. Perhaps one of the most well developed examples relates gravity in asymptotically flat spacetimes in three spacetime dimensions with QFTs that are invariant under the Bondi, van der Burg, Metzner, Sachs \cite{Bondi:1962px,Sachs:1962zza,Sachs:1962wk} ($\mathfrak{bms}_3$) algebra whose non-vanishing commutation relations are given by ($n,m$ are integers)
    \begin{subequations}\label{eq:BMS3}
        \begin{align}
            [\Lt_n,\,\Lt_m]&=(n-m)\,\Lt_{n+m}+\tfrac{c_L}{12}\,n(n^2-1)\,\delta_{n+m,\,0}\\
            [\Lt_n,\,\Mt_m]&=(n-m)\,\Mt_{n+m}+\tfrac{c_M}{12}\,n(n^2-1)\,\delta_{n+m,\,0}\,.
        \end{align}
    \end{subequations}
The generators $\Lt_n$ and $\Mt_n$ are called `superrotations' and `supertranslations', respectively. What makes flat holography such a fertile ground (see e.g. \cite{Barnich:2006av,Barnich:2010eb,Bagchi:2010eg,Barnich:2012xq,Bagchi:2012xr,Afshar:2013vka,Barnich:2015mui,Campoleoni:2015qrh,Bagchi:2016bcd,Riegler:2016hah,Basu:2017aqn}) is that many basic concepts from AdS/CFT can still be applied with slight modifications or even directly obtained from limits of vanishing cosmological constant, $\Lambda\rightarrow0$ \cite{Barnich:2012aw}.

We can now sharpen our main goal, namely to provide non-trivial evidence for the existence of QECs for $\mathfrak{bms}_3$ invariant QFTs exploiting flat holography. 

Our strategy for achieving this is to first derive equations that can be interpreted as saturated versions of such putative inequalities, which we achieve by uniformization methods, reminiscent of conceptually similar methods used in a CFT$_2$ context. We then make use of the limit of vanishing cosmological constant in the gravity dual to the CFT$_2$ obeying QNEC \eqref{eq:AdSQNEC} and show that QECs also exist for $\mathfrak{bms}_3$ invariant QFTs.

\paragraph{Flat space vacuum solutions.}
States in a CFT$_2$ dual to vacuum solutions to three-dimensional Einstein gravity with negative cosmological constant, the so-called Ba{\~n}ados geometries \cite{Banados:1998gg}, saturate QNEC \eqref{eq:AdSQNEC} \cite{Ecker:2019ocp}. Thus, it is natural to expect that geometries corresponding to the flat limit of these geometries provide a good starting point to look for the saturated version of a putative QEC for $\mathfrak{bms}_3$ invariant QFTs. In the following we compute the holographic entanglement entropy (HEE) for these spacetimes and derive equations that can be interpreted as a $\mathfrak{bms}_3$ version of the saturated expression \eqref{eq:AdSQNEC}.


The flat space limit of the Ba{\~n}ados solutions in AdS$_3$ written in retarded Bondi coordinates $0\leq r<\infty$, \mbox{$-\infty<u<\infty$}, $\varphi\sim\varphi+2\pi$ is given by metrics \cite{Barnich:2012rz}
    \begin{equation}\label{eq:FSB}
        \extd s^2= \mathcal{M}(\varphi)\extd u^2-2\extd u\extd r+2\mathcal{N}(u,\varphi)\extd u\extd\varphi+r^2\extd\varphi^2
    \end{equation}
where $2\dot{\mathcal{N}}=\mathcal{M}^\prime$. Here and in what follows dot denotes $\partial_u$ and prime denotes $\partial_\varphi$. 

The solutions \eqref{eq:FSB} include flat space cosmologies (FSC) \cite{Cornalba:2002fi,Cornalba:2003kd} for constant $\mathcal{M}>0$, $\mathcal{N}\neq0$, global flat space for $\mathcal{M}=-1$, $\mathcal{N}=0$, as well as the null orbifold \cite{Horowitz:1990ap,FigueroaO'Farrill:2001nx,Liu:2002ft} for $\mathcal{M}=\mathcal{N}=0$ and $\varphi\rightarrow x$ with $-\infty<x<\infty$. The metric \eqref{eq:FSB} depends on the retarded time $u$ and thus one would have to resort to covariant holographic methods as proposed in \cite{Hijano:2017eii} in contrast to other established methods such as e.g.~\cite{Bagchi:2014iea,Basu:2015evh,Jiang:2017ecm} in order to compute HEE. Alternatively, one can also make use of the fact that the spacetime \eqref{eq:FSB} is locally diffeomorphic to the null orbifold and map a boundary interval in the null orbifold to a boundary interval in \eqref{eq:FSB}. This procedure yields the HEE of \eqref{eq:FSB} from the null orbifold HEE. The same strategy was employed to compute (H)EE for CFT$_2$ states dual to Ba{\~n}ados solutions in AdS$_3$ \cite{Roberts:2012aq,Sheikh-Jabbari:2016znt}. We intend to use the same method and thus need to establish a uniformization map.


\paragraph{Uniformization and Flat Hill's equation.} Consider finite diffeomorphisms
    \begin{subequations}\label{eq:Uniformization}
        \begin{align}
            U&=-2\int\frac{\nu}{\mu^3}\extd\varphi-\frac{2\nu^2}{\mu^2R}\qquad
            x=\int\frac{1}{\mu^2}\extd\varphi+\frac{2\nu}{\mu R}\\
            R&= r\mu^2+2(\nu\mu'-\dot{\nu}\mu)
        \end{align}
    \end{subequations}
with $\dot{\mu}=0$, $\dot{\nu}=\mu'$, where in addition the functions $\mu$ and $\nu$ satisfy the $\mathfrak{bms}_3$ equivalent of Hill's equation \footnote{%
These equations are equivalent to the flat space monodromy equations, Eqs.~(3.10) and (3.13) in \cite{Hijano:2018nhq}.
}
    \begin{equation}\label{eq:iHill}
            \mu''-\frac{\mathcal{M}}{4}\mu=0\qquad\qquad 
            \nu''-\frac{\mathcal{M}}{4}\nu-\frac{\mathcal{N}}{2}\mu=0\,.
    \end{equation}
It should be noted that by replacing $\int\frac{\extd\varphi}{\mu^2}\rightarrow\xi$ and $-2\int\frac{\nu}{\mu^3}\extd\varphi\rightarrow\zeta$ it is obvious that these equations describe precisely how the energy-momentum operators transform under finite $\mathfrak{bms}_3$ transformations \cite{Basu:2015evh,Riegler:2016hah}. 

The diffeomorphism \eqref{eq:Uniformization} with \eqref{eq:iHill} provides a uniformization map, as it locally maps any metric of the form \eqref{eq:FSB} to the null orbifold
    \begin{equation}\label{eq:NOMetric}
        \extd s^2= -2\extd U\extd R+R^2\extd x^2\,.
    \end{equation}
One way of obtaining \eqref{eq:Uniformization} is to make a radial expansion ansatz for the finite diffeomorphism, solve the resulting differential equations order by order and then guess the finite form of the resulting series expansion~\footnote{In order to obtain \eqref{eq:Uniformization} we also chose to define the functions $\mu$ and $\nu$ in a particular way such that they satisfy equations of the form \eqref{eq:iHill}.}.

Since the equations \eqref{eq:iHill} are second order differential equations each of these equations has two linearly independent solutions $\mu_{1,2}$ and $\nu_{1,2}$. A convenient choice of normalization for later purposes turns out to be
    \begin{equation}\label{eq:Normalization}
            \mu_1\mu_2'-\mu_1'\mu_2=1,\,\,
            \mu_1\nu_2'-\mu_1'\nu_2+\nu_1\mu_2'-\nu_1'\mu_2=0.
    \end{equation}    


\paragraph{Uniformization of holographic entanglement entropy.} The EE of a $\mathfrak{bms}_3$ invariant field theory defined on an infinitely long strip at zero temperature with central charges $c_L$ and $c_M$, entangling interval $\Delta x$ and $\Delta U$, as well as the UV cutoffs $\epsilon_x$ and $\epsilon_U$, is given by \cite{Bagchi:2014iea,Jiang:2017ecm}
    \begin{equation}\label{eq:NOHEE}
        S_{\textrm{EE}}=\frac{c_L}{6}\,\log\frac{\Delta x}{\epsilon_x} + \frac{c_M}{6}\,\bigg(\frac{\Delta U}{\Delta x}-\frac{\epsilon_U}{\epsilon_x}\bigg)\,.
    \end{equation}
The dual gravitational description for this expression is given by a Wilson line \cite{Bagchi:2014iea,Basu:2015evh} (or certain geodesics \cite{Jiang:2017ecm}) attached to the boundary of the null orbifold \eqref{eq:NOMetric}. Thus one can obtain EE for a highly excited state that is dual to \eqref{eq:FSB} by applying the diffeomorphism \eqref{eq:Uniformization} to the boundary interval points (and the corresponding cutoffs) of \eqref{eq:NOHEE}.

For large $R$ and taking into account the normalization \eqref{eq:Normalization} the boundary coordinates transform as
    \begin{equation}\label{eq:BdrPointsTrafo}
            U\sim\frac{\mu_1\nu_2-\mu_2\nu_1}{\mu_2^2}\qquad\qquad
            x\sim-\frac{\mu_1}{\mu_2}\,.
    \end{equation}
Applying this diffeomorphism to the initial ($x_i$, $u_i$) and final points ($x_f$, $u_f$) of the intervals $\Delta x=x_f-x_i$ and $\Delta u=u_f-u_i$ \eqref{eq:NOHEE} also means that one has to rescale the UV cutoffs appropriately as
    \begin{equation}\label{eq:UVCutoffTrafo}
            \epsilon_U=\frac{\epsilon_u}{\mu_2^i\mu_2^f}-\epsilon_\varphi\frac{\mu_2^i\nu_2^f+\mu_2^f\nu_2^i}{(\mu_2^i\mu_2^f)^2}\qquad\quad
            \epsilon_x=\frac{\epsilon_\varphi}{\mu_2^i\mu_2^f}
    \end{equation}
where $\epsilon_u$ and $\epsilon_\varphi$ are the UV cutoffs for the coordinates used in \eqref{eq:FSB}. Introducing
    \begin{subequations}
    \label{eq:lalapetz}
        \begin{align}
            \ell_u&:=\mu_1^i\nu_2^f-\mu_1^f\nu_2^i+\mu_2^f\nu_1^i-\mu_2^i\nu_1^f\\
            \ell_\varphi&:=\mu_1^i\mu_2^f-\mu_1^f\mu_2^i\quad\Rightarrow\quad\dot\ell_u = \ell_\varphi^\prime
        \end{align}
    \end{subequations}
establishes the uniformized result for (H)EE
	\begin{empheq}[box=\fbox]{equation}\label{eq:FSBHEE}
        \phantom{\Big(}S_{\textrm{EE}}=\frac{c_L}{6}\log\frac{\ell_
        \varphi}{\epsilon_\varphi} + \frac{c_M}{6}\,\bigg(\frac{\ell_u}{\ell_\varphi}-\frac{\epsilon_u}{\epsilon_\varphi}\bigg)\,.\phantom{\Big)}
	\end{empheq}
This is our first key result. 
\paragraph{Holographic example.}
Inserting the values of $\mathcal{M}$ and $\mathcal{N}$ that correspond to flat space cosmologies, global flat space and the null orbifold \eqref{eq:FSBHEE} reproduces precisely the known expressions in the literature. For example, setting $\mathcal{M}=-1$ and $\mathcal{N}=0$, solving Flat Hill's equation \eqref{eq:iHill} subject to the normalizations \eqref{eq:Normalization} yields the solutions
$\mu_1=\alpha\cos(\frac{\varphi}{2})$, $\nu_1=\gamma\cos(\frac{\varphi}{2})+u\,\mu_1'$,
$\mu_2=\frac{2}{\alpha}\sin(\frac{\varphi}{2})$, $\nu_2=-\frac{2\gamma}{\alpha^2}\sin(\frac{\varphi}{2})+u\,\mu_2'$
for some constants $\alpha$ and $\gamma$. Plugging these solutions into \eqref{eq:FSBHEE} with \eqref{eq:lalapetz} gives
\eq{
S_{\textrm{EE}}=\frac{c_L}{6}\,\log\bigg[\frac{2\sin\big(\frac{\Delta\varphi}{2}\big)}{\epsilon_\varphi}\bigg] + \frac{c_M}{6}\bigg(\frac{\Delta u}{2} \cot\Big(\frac{\Delta\varphi}{2}\Big)-\frac{\epsilon_u}{\epsilon_\varphi}\bigg)
}{eq:HE1}
in agreement with HEE of global flat space \cite{Bagchi:2014iea}\footnote{%
When comparing \eqref{eq:HE1} with Eq.~(20) in \cite{Bagchi:2014iea} we need to set $\epsilon_\varphi=1$, $\epsilon_u=0$, $\Delta\varphi=\xi_{12}$, $\Delta u=\rho_{12}$ and $L=2\pi$.
}.

For later purposes we define separately the contributions that arise when either $c_L$ or $c_M$ vanish,
\eq{
S_L:=\frac{c_L}{6}\,\log\frac{\ell_\varphi}{\epsilon_\varphi} \qquad
S_M:=\frac{c_M}{6}\,\bigg(\frac{\ell_u}{\ell_\varphi}-\frac{\epsilon_u}{\epsilon_\varphi}\bigg) 
}{eq:angelinajolie}
in terms of which \eqref{eq:FSBHEE} reads $S_{\textrm{EE}}=S_L+S_M$.

\paragraph{Entanglement entropy and BMS transformations.} 
One particularly useful byproduct of the uniformization procedure above are the relations \eqref{eq:BdrPointsTrafo} and \eqref{eq:UVCutoffTrafo} that give a precise way to determine how the EE of a $\mathfrak{bms}_3$ invariant QFT on an infinitely long strip transforms under finite BMS transformations. After rewriting these equations one can directly determine the transformation behavior of $S_L$ and $S_M$ under infinitesimal supertranslations and superrotations that generate the $\mathfrak{bms}_3$ algebra.

For that purpose take finite BMS transformations
    \begin{equation}
        x\rightarrow\xi(\varphi),\qquad\quad U\rightarrow \zeta(u,\,\varphi)
    \end{equation}
where $\dot\zeta=\xi^\prime$. These transformations map a set of coordinates $(x,\,U)$ to a new set $(\varphi,\,u)$. For a given set of entangling intervals $\Delta u$ and $\Delta\varphi$ this change of coordinates affects the UV cutoffs $\epsilon_x$ and $\epsilon_U$ as
    \begin{subequations}
        \begin{align}
            \epsilon_x&=\frac{\epsilon_\varphi}{2}\sqrt{\xi'_i+\dot{\zeta}_i}\sqrt{\xi'_f+\dot{\zeta}_f}\\
            \epsilon_U&=\epsilon_u\frac{\epsilon_x}{\epsilon_\varphi}+\frac{\epsilon_\varphi^2}{4\epsilon_x}\left(\zeta'_i(\xi'_f+\dot{\zeta}_f)+\zeta'_f(\xi'_i+\dot{\zeta}_i)\right).
        \end{align}
    \end{subequations}
Here $\xi_{i/f}\equiv\xi(\varphi_{i/f})$, $\zeta_{i/f}\equiv\zeta(u_{i/f},\varphi_{i/f})$, and $(u_i,\varphi_i)$ and $(u_f,\varphi_f)$ denote the initial and final endpoints of the entangling interval, respectively.

The EE \eqref{eq:NOHEE} transforms accordingly as
    \begin{subequations}\label{eq:EEFiniteBMSTrafos}
    \begin{align}
        S_L&=\frac{c_L}{6}\,\log\frac{2(\xi_f-\xi_i)}{\epsilon_\varphi\sqrt{\xi'_i+\dot{\zeta}_i}\sqrt{\xi'_f+\dot{\zeta}_f}}\\
        S_M&=\frac{c_M}{6}\,\bigg(\frac{\zeta_f-\zeta_i}{\xi_f-\xi_i}-\frac{\epsilon_u}{\epsilon_\varphi}-\frac{\zeta'_f}{\xi'_f+\dot{\zeta}_f}-\frac{\zeta'_i}{\xi'_i+\dot{\zeta}_i}\bigg)
    \end{align}
    \end{subequations} 
where we used again the definitions $S_L$ and $S_M$, see \eqref{eq:angelinajolie}.

From this transformation behavior it is straightforward to determine that under infinitesimal transformations comprising superrotations generated by $\sigma$ and supertranslations generated by $\eta$,
    \begin{equation}
        \xi(\varphi)=\varphi+\sigma(\varphi)\qquad\quad \zeta(u,\varphi)=\eta(\varphi)+u\,\xi'(\varphi)
    \end{equation}
the EE transforms as
	\begin{empheq}[box=\fbox]{equation}\label{eq:EEtrafo}
        \delta S_L = S_L'\sigma-\frac{c_L}{12}\sigma'\qquad \delta S_M = S_M'\sigma + \dot{S}_M\zeta-\frac{c_M}{12}\zeta'\,.
	\end{empheq}
These transformation equations are our second key result. The first equality shows that $S_L$ transforms like an anomalous weight-0 scalar in a chiral half of a CFT$_2$ (see e.g.~\cite{Blumenhagen}) under superrotations and is inert under supertranslations. The second equality shows that $S_M$ transforms non-trivially both under superrotations (again with weight-0) and supertranslations.  
    
On a sidenote, in analogy to the CFT$_2$ case (see e.g.~\cite{deBoer:2016pqk,Vaknin:2017yiz,Caputa:2017yrh}) the result \eqref{eq:EEFiniteBMSTrafos} also implies that $S_{L/M}$ satisfy $\mathfrak{bms}_3$ versions of Liouville's equation \cite{Barnich:2012rz},
    \begin{subequations}
        \begin{align}
            \epsilon_\varphi^2\partial_{\varphi_1}\partial_{\varphi_2}S_L & =\frac{c_L}{6}e^{-\frac{12}{c_L}S_L}\\
            \epsilon_\varphi^2\partial_{\varphi_1}\partial_{\varphi_2}S_M & = 2\Big(S_M+\frac{c_M}{6}\frac{\epsilon_u}{\epsilon_\varphi}\Big)e^{-\frac{12}{c_L}S_L}\,.
        \end{align}
    \end{subequations}

\paragraph{Saturation equations.} In similar spirit to QNEC \eqref{eq:AdSQNEC} a natural ansatz for a saturated QEC in $\mathfrak{bms}_3$ invariant QFTs is to linearly combine all possible second derivatives and quadratic powers of first derivatives of EE \eqref{eq:FSBHEE} [or equivalently \eqref{eq:angelinajolie}] so that they reproduce the expectation values of the energy-momentum operators in this QFT. For $c_L\neq 0 \neq c_M$, there is a unique combination:
	\begin{empheq}[box=\fbox]{equation}\label{eq:SaturationEquations}
	\begin{split}
	\phantom{\Big(}
	 2\pi\langle\mathcal{T}_{M}\rangle&=\dot{S}_M'+\frac{6}{c_M}\,\dot{S}_M^2 \\
           2\pi\langle\mathcal{T}_{L}\rangle&= S_L''+\frac{6}{c_L}\,S_L'^2+S_M''+\frac{12}{c_L}\,S_L^\prime S_M^\prime
	\phantom{\Big(}
           \end{split}
	\end{empheq}
The derivatives in \eqref{eq:SaturationEquations} are understood to be taken with respect to one of the boundary endpoints, i.e.~either ($x_i,u_i$) or ($x_f,u_f$). The expectation values of the energy-momentum operators in terms of the charges $\mathcal{M}$ and $\mathcal{N}$ are (see e.g. \cite{Fareghbal:2013ifa})
    \begin{equation}\label{eq:EMTensorAndCharges}
        2\pi\langle\mathcal{T}_{M}\rangle=\frac{c_M}{24}\,\mathcal{M}\qquad 2\pi\langle\mathcal{T}_{L}\rangle=\frac{c_L}{24}\, \mathcal{M}+\frac{c_M}{12}\, \mathcal{N}\,.
    \end{equation}
The saturation equations \eqref{eq:SaturationEquations} are our third key result. They suggest that QECs in $\mathfrak{bms}_3$ invariant QFTs, if they exist, are inequality versions of \eqref{eq:SaturationEquations}.   
    
The saturation equations \eqref{eq:SaturationEquations} are related to the $\mathfrak{bms}_3$ equivalent of Hill's equation \eqref{eq:iHill} by the re-definitions
\mbox{$\mu=\exp[\frac{6}{c_L}S_L]$} and $\nu=\frac{6}{c_M}\,S_M\,\mu$.
This provides a cross check on the validity of the saturation equations \eqref{eq:SaturationEquations}.


\paragraph{QECs as flat space limit of QNEC.} Two-dimensional CFTs and their gravity duals in AdS$_3$ are extremely well-studied, which is why on the gravity side the limit of vanishing cosmological constant $\Lambda = -\epsilon^2 \to 0$ was used extensively in the past in order to gain a better understanding of $\mathfrak{bms}_3$ invariant QFTs. See for example \cite{Bagchi:2012cy,Barnich:2012aw,Krishnan:2013wta,Riegler:2014bia,Fareghbal:2014qga,Campoleoni:2016vsh,Hijano:2019qmi}. 

Following this philosophy we take now the flat space limit of \eqref{eq:AdSQNEC}. In a CFT$_2$ EE splits into two chiral parts as $S_{\textrm{EE}}=S^++S^-$, where
    \begin{equation}
        S^\pm=\frac{c^\pm}{12}\,\log\Big(\frac{\ell^\pm}{\delta^\pm}\Big)^2\,.
    \end{equation}
Here $c^\pm>0$ denote the central charges of the CFT and $\delta^\pm$ the respective UV cutoffs. The functions $\ell^\pm$ only depend on $x^\pm$, respectively, and encode the entangling interval. See e.g.~\cite{Sheikh-Jabbari:2016znt} for more details on the precise form of $\ell^\pm$. The central charges, UV cutoffs, entangling intervals and the derivatives appearing in \eqref{eq:AdSQNEC} are related to their flat space counter parts via
    \begin{subequations}
        \begin{align}
            c^\pm&=\tfrac{1}{2}\big(\frac{c_M}{\epsilon}\pm c_L\big) &
            \delta^\pm &=\epsilon_\varphi\pm\epsilon\,\epsilon_u \\
            \ell^\pm&=\epsilon\ell_u\pm\ell_\varphi & \partial_\pm&=\tfrac{1}{2}\Big(\frac{\partial_u}{\epsilon}\pm\partial_\varphi\Big)\,.
        \end{align}
    \end{subequations}
The chiral conditions $\partial_\mp S^\pm=0$ imply $\frac{1}{c_M}\dot S_M=\frac{1}{c_L}S_L^\prime$ as well as $\frac{1}{c_L}\dot S_L=\frac{\epsilon^2}{c_M} S_M^\prime$. It can be readily checked that the combination~\footnote{See e.g. \cite{Riegler:2014bia,Jiang:2017ecm} for an in-depth explanation why the difference and not the sum of the two chiral CFT halves has to be considered for the flat space limit.} $\lim_{\epsilon\rightarrow0}\left(S^+-S^-\right)$ precisely reproduces \eqref{eq:FSBHEE}, which provides an additional cross check of our earlier results.

The next step is to rewrite the QNEC inequalities \eqref{eq:AdSQNEC} as
    \eq{
     2\pi\langle\mathcal{T}_{\pm\pm}\rangle=\partial_\pm^2S^\pm + \frac{6}{c^\pm} \big(\partial_\pm S^\pm\big)^2 + c^\pm a^\pm
    }{eq:CFTQNECLimit}
where $a^\pm\geq0$. The components of the CFT energy-momentum tensor $\mathcal{T}_{\pm\pm}$ are related to their $\mathfrak{bms}_3$ counterpart via \cite{Bagchi:2013bga,Fareghbal:2013ifa}
    \begin{equation}\label{eq:CFTEMLimit}
        \mathcal{T}_{M}=\left(\mathcal{T}_{++}+\mathcal{T}_{--}\right)\epsilon\qquad \mathcal{T}_{L}=\mathcal{T}_{++}-\mathcal{T}_{--}\,.
    \end{equation}
Linearly combining the right hand sides of \eqref{eq:CFTQNECLimit} as in \eqref{eq:CFTEMLimit} and taking the limit $\epsilon\rightarrow0$ requires $a$ and $b$ to scale with $\epsilon$ as
    \begin{equation}\label{eq:AandBLimit}
        a^\pm = \tfrac{1}{2}\,\big(\beta\pm\epsilon\alpha\big)
    \end{equation}
in order to obtain a finite result, namely
    \begin{subequations}\label{eq:BMSNonSaturation}
        \begin{align}
            2\pi\langle\mathcal{T}_{M}\rangle&=\dot{S}_M'+\frac{6}{c_M}\,\dot{S}_M^2+\frac{c_M}{2}\,\beta \label{eq:BMSNonSaturation1}\\
           2\pi\langle\mathcal{T}_{L}\rangle&= S_L''+\frac{6}{c_L}\,S_L'^2+\frac{c_L}{2}\,\beta\nonumber\\
           &\quad+S_M''+\frac{12}{c_L}\,S_L^\prime S_M^\prime +\frac{c_M}{2}\,\alpha\,. \label{eq:BMSNonSaturation2}
        \end{align}
    \end{subequations}
From \eqref{eq:AandBLimit} and the conditions $a^\pm\geq0$ it is evident that also $\beta\geq0$ has to hold in the limit $\epsilon\rightarrow0$. The sign of $\alpha=(a^+-a^-)/\epsilon$, however, is in general not fixed in this limit. In fact there is even a very simple example showing that $\alpha$ can be either positive or negative.

Adding matter is one way to drive QNEC out of saturation in a CFT. For holographic CFTs this can be achieved for example by considering shock waves as in e.g.~\cite{Khandker:2018xls}. A special case of such shock waves is described by the following metric in Poincar{\'e} coordinates,
    \begin{equation}\label{eq:PPShockWavemetric}
        \frac{\extd s^2}{\ell^2}= \frac{\extd z^2+\extd x^+\extd x^-}{z^2}+h_{\pm\pm}(x^\pm,z)\left(\extd x^\pm\right)^2\,.
    \end{equation}
In the dual CFT either of these shock waves result in a chiral non-saturation of QNEC. After computing EE holographically using the metric \eqref{eq:PPShockWavemetric} and taking the appropriate derivatives one obtains \eqref{eq:CFTQNECLimit} with either $a^+\geq0$ and $a^-=0$ for $h_{++}\neq0$ or $a^+=0$ and $a^-\neq0$ for $h_{--}\neq0$. This simple example shows that $\alpha$ in general can be either positive or negative in the limit of vanishing cosmological constant.


\paragraph{Flat space QECs.} Taking into account the results from the previous section and the three key results, we propose the following inequalities to hold for $\mathfrak{bms}_3$ invariant QFTs with central charges $c_M$ and $c_L$. 	

Theories with $c_M=0$ are essentially a chiral half of a CFT$_2$ and hence assuming unitarity, in particular $c_L>0$, the following inequality has to be obeyed
	\begin{empheq}[box=\fbox]{equation}\label{eq:FlatQNEC1}
	\phantom{\Big(}2\pi\langle\mathcal{T}_{L}\rangle\geq S_L''+\frac{6}{c_L}\,{S_L'}^{\!2}\phantom{\Big).}
	\end{empheq}
This inequality resembles the QNEC inequality \eqref{eq:AdSQNEC}.

For generic theories with $c_M\neq 0$ one has instead
	\begin{empheq}[box=\fbox]{equation}\label{eq:FlatQNEC2}
		\phantom{\Big(}2\pi\langle\mathcal{T}_{M}\rangle\geq\dot{S}_M'+\frac{6}{c_M}\,\dot{S}_M^{\,2}\,.\phantom{\Big)}
	\end{empheq}
The two QECs above are our main result.

\paragraph{Summary and further developments.} Our derivation of the uniformization result for entanglement entropy \eqref{eq:FSBHEE}, its transformation properties \eqref{eq:EEtrafo}, the saturation equations \eqref{eq:SaturationEquations} and our arguments leading to the quantum energy conditions \eqref{eq:FlatQNEC1}, \eqref{eq:FlatQNEC2} were all based on holographic considerations and on calculations done on the gravity side. The main reason why we are sure that the same results must hold on the field theory side is that the key ingredient in all our considerations were the symmetries \eqref{eq:BMS3}, that by definition govern any $\mathfrak{bms}_3$ invariant QFT. Still, it could be worthwhile to rederive all our statements intrinsically from a field theoretic perspective, using e.g.~properties of Galilean conformal field theories \cite{Bagchi:2009my,Bagchi:2010eg}. This will be particularly interesting for appearances of $\mathfrak{bms}_3$ symmetries beyond flat space holography such as tensionless bosonic string theory, see e.g.~\cite{Bagchi:2015nca,Bagchi:2016yyf,Bagchi:2019cay}.

The quantum energy conditions \eqref{eq:FlatQNEC1}, \eqref{eq:FlatQNEC2} are expected to play a similar role for the understanding of $\mathfrak{bms}_3$ invariant QFTs as the quantum null energy condition for Lorentz-invariant QFTs. It would be rewarding to translate relativistic results based on the quantum null energy condition to corresponding results in $\mathfrak{bms}_3$ invariant QFTs.

Finally, what we have provided was merely the first example of quantum energy conditions in a specific class of non-Lorentz-invariant field theories. However, there are very likely more examples to be discovered. Candidate theories that plausibly will exhibit such inequalities are warped CFTs \cite{Detournay:2012pc}, which share relevant features with two-dimensional CFTs and $\mathfrak{bms}_3$ invariant QFTs, in particular an infinite set of symmetries.

\acknowledgments

The authors thank Tarek Anous, Luis Apolo, Arjun Bagchi, Aritra Banerjee, Rudranil Basu, Pawel Caputa, Soumangsu Chakraborty, Pankaj Chaturvedi, Geoffrey Comp\`{e}re, St{\'e}phane Detournay, Christian Ecker, Masahiro Nozaki, Matthew M. Roberts, Shahin Sheikh-Jabbari, Wei Song, Julian Sonner, Philipp Stanzer, Tadashi Takayanagi, Quentin Vandermiers and Wilke van der Schee for enlightening discussions. 
DG was supported by the FWF projects P30822 and P28751. The research of MR is partially supported by the ERC Advanced Grant ``High-Spin-Grav'' and by FNRS-Belgium (convention FRFC PDR T.1025.14 and  convention IISN 4.4503.15). PP is funded by the Junior Research Fellowship Programme from ESI Vienna, and  acknowledges the hospitality of TU Wien. Part of this work was discussed during the Programme and Workshop ``Higher spins and holography'' in March/April 2019 at the Erwin Schr\"odinger International Institute for Mathematics and Physics (ESI), and DG and PP thank ESI for hosting us. 
\bibliography{Bibliography}

\end{document}